# Orbital angular momentum mode selection by rotationally symmetric superposition of chiral states with application to electron vortex beams

Yuanjie Yang[1,2*], G. Thirunavukkarasu[1], M. Babiker[1], Jun Yuan[1*]

[1]Department of Physics, University of York, Heslington, York YO10 5DD, UK

[2]School of Astronautics & Aeronautics, University of Electronic Science and Technology of China, Chengdu 611731, China

A general orbital angular momentum (OAM) mode selection principle is put forward involving the rotationally symmetric superposition of chiral states. This principle is not only capable of explaining the operation of spiral zone plate holograms and suggesting that naturally occurring rotationally symmetric patterns could be inadvertent sources of vortex beams, but more importantly, it enables the systematic and flexible generation of structured OAM waves in general. This is demonstrated both experimentally and theoretically in the context of electron vortex beams using rotationally symmetric binary amplitude chiral sieve masks.

Vortices are common to all wave phenomena [1], including tornadoes at the large scale and superfluid helium at the small scale. The prototypical example of these is the optical vortex beam [2], followed recently by the electron vortex beam [3-5] and there are early studies of neutron [6] and atom vortex beams [7]. Vortex beams are characterized by a phase singularity described by an $\exp(i\ell\theta)$ azimuthal phase factor, where $\theta$ is the azimuthal angle and $\ell$ stands for the winding number and is also called the topological charge. Vortex beams are of special interest because of their quantized OAM of $\ell\hbar$ per particle [1] and have led to various applications in the contexts of super-resolution microscopy [8], nano-manipulation [9], astronomy [10] and crystallography [11]. In contrast to the intrinsic (spin) angular momentum, OAM can be very large and, as such, vortex beams can lead to new physics [12, 13] and have potential in multiplex free-space communication [14] and quantum information [15].

The establishment of a toolbox for flexible vortex beam generation is the key to the development of science and technology involving vortex beams. If the vortex beam wavefunctions are known exactly, they can be generated either by direct phase manipulation (or wavefront shaping) [16] using, for example, spiral phase plates [3,6,17], spin-to-orbital angular momentum convertors [18], and by phase encoding techniques through diffraction involving computer-generated holograms (CGHs) [19], including fork grating [4,5] and spiral zone plates [20-22]. Other techniques require case by case analysis to identify the nature of the vortex beams produced, as in the case involving the Aharonov–Bohm effect experienced by a charged particle in a suitable magnetic field [23,24]. More recent methods make use of photon sieves [25-27] and Vogel spiral arrays [28, 29] as diffractive elements for vortex beam generation.

Here we put forward a general principle for vortex beam generation in rotationally symmetric systems. We examine the role of rotational symmetry, not only for the purpose of generating a specific vortex wavefunction, but also for the essential symmetry elements that must be possessed by any vortex-related state by virtue of its characteristic azimuthal phase factor $\exp(i\ell\theta)$. The principle would enable us to understand, in a novel deconstructive manner, the generation of individual pure vortex beams using a diverse range of existing rotationally symmetric optical elements, such as spiral diffractive holograms (spiral zone plates) [20-22], 2D chiral cam-shaped objects [30] and plasmonic vortex lens [31]. Furthermore, the study of rotational symmetry in the vortex context would provide a useful guide for exploring a broader range of vortex-related beams, including those with complex vortex characters, such as vortex modes which consist of a mixture of concentric vortex states each with a different OAM content.

The principle in question emerges from the following analysis. Consider the scenario in which $m$ identically monochromatic waves are equally spaced in the azimuthal angular domain. Let $u\left(\rho, z, [\theta + \frac{2\pi s}{m}]\right)$ be the complex amplitude of the $s^{th}$ wave in cylindrical polar coordinates. Then the total field due to the rotationally symmetric superposition of $m$ such waves is the sum:

$$\psi(\rho, z, \theta) = \sum_{s=0}^{m-1} u\left(\rho, z, [\theta + \frac{2\pi s}{m}]\right) \quad (1)$$

The individual amplitude function $u(\rho, z, \theta)$ can be expressed in terms of any complete orthonormal basis set such as the set of Laguerre-Gaussian (LG) modes [15]. A useful alternative would be the complete set of the Fourier transforms of the truncated Bessel (FT-TBB) functions [32]. We consider here the use of the LG set and write

$$u(\rho, z, \theta) \propto \sum_{p=0}^{\infty} \sum_{\ell=-\infty}^{\infty} c_{p,\ell} \varphi_{p,\ell}(\rho, z, \theta) \quad (2)$$

where $\varphi_{p,\ell}(\rho, z, \theta)$ denotes a LG mode of radial and azimuthal indices $(p, \ell)$ and the expansion coefficients $c_{p,\ell} = \iint u(\rho,z,\theta)\varphi^*_{p,\ell}(\rho,z,\theta)\rho d\rho d\theta$ are the overlap integrals. Substituting Eq. (2) into Eq. (1) we have

$$\psi(\rho,z,\theta) = \sum_{s=0}^{m-1}\sum_{p=0}^{\infty}\sum_{\ell=-\infty}^{\infty} c_{p,\ell}\varphi_{p,\ell}\left(\rho,z,[\theta+\frac{2\pi s}{m}]\right) \quad (3)$$

The LG modes have the form $\varphi_{p,\ell}(\rho,z,\theta) = A_{p,\ell}(\rho,z)\exp(i\ell\theta)$ where $A_{p,\ell}(\rho,z)$'s are the mode amplitude functions. Once this form of $\varphi_{p,\ell}(\rho,z,\theta)$ is inserted in Eq.(3), it becomes clear that the finite series summation over $s$ is just a geometrical series summation which yields:

$$S = \sum_{s=0}^{m-1}\exp\left(i\ell\frac{2\pi s}{m}\right) = \frac{1-\exp\left(i\ell\frac{2\pi m}{m}\right)}{1-\exp\left(i\ell\frac{2\pi}{m}\right)}$$
$$= \begin{cases} m, \ell = Mm \\ 0, \ell \neq Mm \end{cases} \quad (4)$$

where M is an integer, namely M = $0, \pm 1, \pm 2, \dots$. As a result, Eq. (3) gives:

$$\psi(\rho,\theta,z) = \begin{cases} m\sum_{\ell}\sum_{p=0}^{\infty} c_{p,\ell}A_{p,\ell}(\rho,z)\exp(i\ell\theta), \ell = Mm \\ 0, \ell \neq Mm \end{cases} \quad (5)$$

Recall that $m$ is the number of equally spaced identically monochromatic waves and M is an integer. Equation (5) embodies what we call *the principle of rotationally symmetric superposition of chiral states* and has two interesting consequences:

 (i) If the initial waves are pure vortex beam states with well-defined OAM, then Eq. (5) is reduced to

$$\psi(\rho,\theta,z) = \begin{cases} mu(\rho,z,\theta), \ell = Mm \\ 0, \ell \neq Mm \end{cases} \quad (6)$$

which means that the superposition of LG modes is non-vanishing only when the topological charge $\ell$ is a multiple of $m$, the number of waves.

(ii) If the initial waves are not vortex states with pure OAM modes, then Eq.(5) shows that the superposition of $m$ such waves can produce vortex beams with OAM modes $\ell = 0, \pm m, \pm 2m, \dots$. If a specific single OAM mode is required out of the multiple OAM harmonics allowed by the rotational symmetry of the system, two further 'filtering' processes would be needed to achieve that goal: (1) the inclusion of chiral symmetry to break the mirror symmetry; and (2) the selection of the particular OAM mode.

As a first illustration of these OAM 'combing' and 'filtering' processes we consider how to generate a pure vortex beam with a specific topological charge. In this example the wavelets are due to diffracted beams from plane waves passing through a rotationally symmetric arrangement of pinholes, as shown in Fig.1. The result of the superposition is observed in a plane distance $z$ away. Each of the wavelets contains a geometrical phase factor that can be manipulated by adjusting the geometry and positions of the pinholes. Figure 1(a) shows the simplest case of a 5-fold rotationally symmetric *achiral* mask consisting of only five pinholes. When the mask is illuminated by a plane wave, it leads to the intensity and phase patterns at the observation $z$ plane as shown in Fig.1(b) and Fig.1(c), respectively.

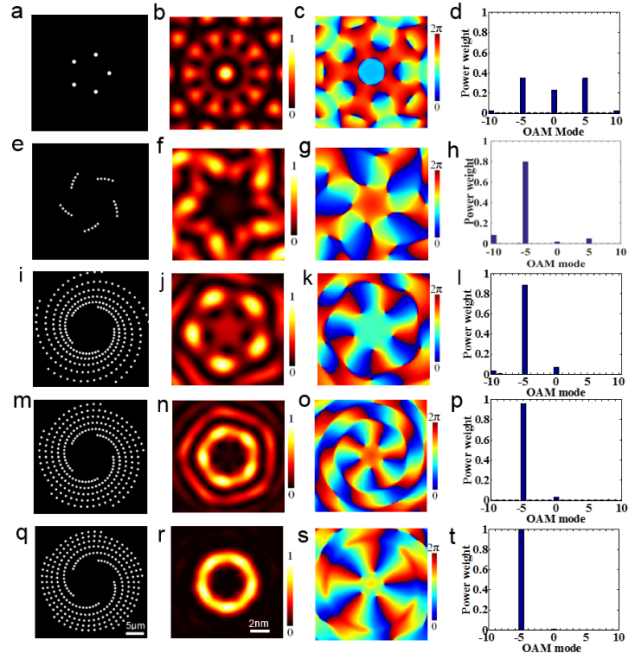

FIG. 1 (color online). Management of the OAM modes using rotationally symmetric masks constructed from different motifs. (a) five-pinhole mask; (b) the intensity pattern and (c) phase pattern at defocus $\Delta f = -33.6$ μm, and (d) the corresponding power spectra in the OAM mode $\ell$. (e)-(h) based on 5 short pinhole-curves (motifs). (i-l) based on Logarithmic spirals; (m-p) based on Archimedean spirals. (q-t) based on the Fermat spirals. Here, the expansion area of the complex fields which were decomposed into the LG basis set was 10nm×10nm.

The theoretical relative power of the individual modes emerging on illumination is obtainable by the decomposition of the complex field in terms of the LG basis set [15]. The power spectrum corresponding to the mask in Fig. 1(a) is displayed in Fig.1(d) which shows that the five-pinhole mask converts a plane wave into a set of OAM modes with $\ell = 0, \pm 5, \pm 10$, which is consistent with the result in Eq. (5). The symmetric distribution of the power spectrum is indicative of the mirror symmetry of the pinhole arrangement.

In order to generate vortex beams with chiral phase structures, we have to deal with a 'chiral' rotationally symmetric pinhole array mask. The first example of such a mask is shown in Fig.1(e) consisting of 5 short pinhole 'motifs' each with several pinholes arranged as shown. For an effective vortex beam production, individual motifs in the designed mask would typically consist of many pinholes distributed along a number of spirals. Spiral pinhole motifs are particularly attractive as the corresponding masks have a clear handedness. The corresponding OAM mode spectrum shown in Fig. 1(h) is seen to be dominated by the $\ell = -5$ mode. The power weights of the other modes are very low, as desired. This OAM selection ability is a major feature of the principle we have introduced above.

To explore this OAM-dependent on-axis 'focusing' effect of the chiral multiple-pin-hole mask, we have considered pinhole masks in which the basic unit motifs are in the form of well-known spirals. The last three rows in Fig.1 show the cases of spiral-pinhole masks involving Logarithmic ($\rho \propto a\exp(b\theta)$), Archimedean ($\rho \propto (a+b\theta)$) and Fermat spiral motifs. In the case of the Fermat spiral, N pinholes are distributed along each spiral motif according to $\alpha_n = 2\pi n/N$ and $r_n = (r_0^2 + \ell z\lambda\alpha_n/\pi)^{1/2}$, where $\lambda$ is the wavelength of the incident wave, $z$ is the observation plane, and $r_0$ is the coordinate of the first pinhole from the centre [33]. The results of the OAM mode analysis for this case are shown in Figs. 1(l), 1(p) and 1(t). These results indicate that the purity of the OAM of the vortex modes generated near the beam axis can be progressively tuned by using masks based on different kinds of spirals.

Although the positions of the center of the set of holes in a single Fermat spiral are designed to produce an on-axis angular dependent geometrical phase structure consistent with the azimuthal phase ramp of an OAM beam of winding number $\ell$ [33], our study shows that the contributions of the waves from the set of pinholes do not produce the desired OAM mode solely, and the vortex beam produced by a single Fermat spiral has a wide OAM spectrum. The rotationally symmetric arrangement of $\ell$ Fermat spiral motifs reinforces the strength of the allowed OAM modes generated, while 'filtering' out the neighboring symmetry-incompatible OAM modes, resulting in the purist on-axis OAM mode, as shown by the OAM spectrum in Fig. 1(t). The resulting spiral multi-pinhole mask shown in Fig. 1(q) may be considered as a simplified version of the spiral diffraction grating mask designed to produce a pure vortex beam of winding number $\ell$ at the observation plane. Unlike the spiral CGH mask, the multi-pinhole masks are easy to manufacture and are mechanically more robust since our deconstructive analysis allows us to retain the essential features required for the pure vortex beam generation while achieving similar vortex beam conversion efficiency.

However, the true power of our 'deconstructed approach' lies in the considerable degree of freedom afforded by the systematic and rational design of more realistic and complex vortex beam approximates that, to date, are difficult to foresee and achieve, neither by the traditional CGH approach nor by those based on other ad hoc bases [28, 29].

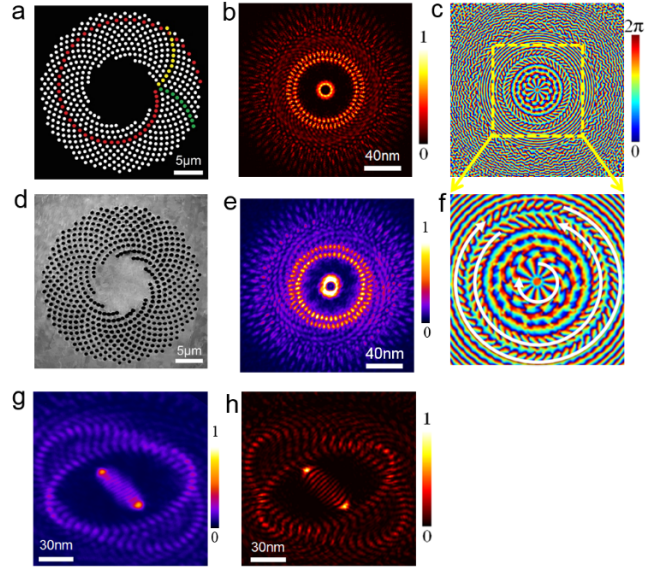

FIG. 2 (color online). Generation of concentric electron vortex beams and the measurement of their topological charges. (a) Simulation of a structured electron sieve. (b) and (c) the corresponding simulations of the intensity (b) and phase (c) at defocus $\Delta f = -74.6$ μm. (d) The scanning electron microscope image of the electron sieve. (e) The experimental results of the intensity patterns. (f) a zoom-in view of the center part of (c). The white curves in (f) indicate increasing direction of phase. (g) and (h) the experimental results (g) and the corresponding simulation (h) of the intensity pattern after astigmatic transformation.

Finally, as a further illustration of the new insight gained from our analysis, we consider the compact pinhole mask shown in Fig. 2 (a). This kind of mask design is suitable for the production of vortex beams with more complicated spatial structures. The mask in Fig. 2(a) is designed to have 11 Fermat spirals (red), each of which is the same as that that given in Fig.1(q). Due to the degeneracy, two additional spirals with repetition of 44 (yellow spirals) and 55 (green spirals) can be identified. Moreover, the handedness of the yellow spirals is opposite to those of the red and green spirals. According to the above analysis, we expect each of such regular spiral arrangements to support an OAM mode. Unlike the vortex beams generated by the spiral CGH masks where different OAM modes are 'focused' at different on-axis positions while other OAM modes exist as a complex mixture in the background, our masks generate a beam with three bright rings corresponds to the three sets of spiral pattern in the mask (Fig. 2(b)), with the outer two rings having 44 and 55 bright dots respectively. The simulated phase map of the diffracted beams shown in Figs. 2(c) and 2(f) suggests that the total phase change for each over a complete revolution is by $-11 \times 2\pi$ around the center, by $44 \times 2\pi$ and $-55 \times 2\pi$ in the corresponding regions of two necklace-like beams in the clockwise direction. This shows that the 'sieve' mask has generated a compound vortex beam containing simultaneously three co-propagating vortex beams with topological charges $\ell$ = -11, 44 and -55, respectively.

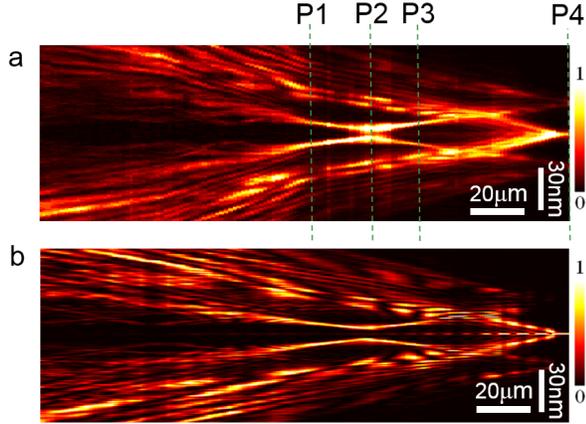

FIG. 3 (color online). The propagation of vortex beams generated by electron sieves. (a) Experimental results of the intensity pattern in the y-z plane. (b) The corresponding result of simulation. The experimental result in (a) was reconstructed from 140 slices of the x-y plane intensity pattern recorded in the experiment. P1-P4 represent four transverse planes perpendicular to the propagation axis. P2 corresponds to the intensity patterns shown in Fig.2e, and P4 denotes the focal plane of the condenser lens.

Note that our analysis is general and is therefore applicable to any vortex states that can be described by a scalar wave equation, such as optical vortex beams, electron vortex beams and other matter vortex beams.

We now provide a description of the experimental work we have carried out to test the above theoretical predictions applicable in the context of electron vortex beams. The physical mask presented in Fig. 2(d) was created from a 1μm thick Pt foil and the radius of each pinhole $a$ = 300 nm. We illuminated the mask with a relatively coherent electron beam in a JEOL 2200FS TEM operating at 200 kV, which corresponds to an electron de Broglie wavelength of $\lambda$ =2.5 pm. The mask was inserted in the condenser aperture of the electron microscope. Instead of observing the pattern directly, the condenser lens was turned on and the observation is conducted at the corresponding defocus distance from the focal plane of the condenser lens which has a focal length of about 15 mm. Figure 2(e) displays the experimental intensity pattern bearing three rings, in good agreement with the simulated intensity pattern (Fig.2(b)). It is well known that the vortex modes can be astigmatically transformed into Hermite-Gauss-like modes, with the number of dark stripes indicating the value of the OAM carried by the vortex [34]. The experimental intensity pattern (Fig.2(g)) and the corresponding simulation result (Fig.2(h)) of the vortex beam after the astigmatic transformation show that indeed there are 11 dark stripes in the center, with respect to the tilt direction, confirming that the inner ring of the emerging beam is a vortex beam with topological charge $\ell$ = −11. Furthermore, both Fig.2(g) and Fig.2(h) also exhibit two ellipses tilted in different directions, consistent with the fact that the topological charges of the two outer rings of the intensity pattern (Figs.2(b) and (e)) do indeed have opposite signs [34].

We can further explore the characteristics of the electron vortex state consisting of three co-propagating OAM modes by following the evolution of the intensity distribution near the focal plane (P4 in Fig.3) of the condenser lens. The experimental observation displayed together with the simulated normalized intensity distribution in the y-z plane in Fig.3 constitute clear agreement, including the confirmation that the Fermat spirals have led to an OAM mode of topological charge of magnitude 11 which has its beam waist at the observation plane Δf = -74.6 μm. (P2 in Fig.3). The two other spirals in the sieve mask are non-Fermat-like and hence have no clear beam waist at the plane P2. The different focusing behaviors of the three vortex rings also mean they have different Gouy phase change near the observation plane. This is consistent with the different stages of the astigmatic transformation of the three OAM rings seen in Fig. 2(g, h). The Supplementary Movie indicated in Ref.[35] shows that these outer two necklace-like rings have opposite senses of rotation and each maintains its fixed number of pearls when they propagate from plane P1 to plane P3, demonstrating the correct identification of the helicity of the vortex mode.

We reiterate that our general design approach and the results we have obtained with electron sieve masks are also applicable to other matter waves as well as optical vortex beams. We would like to highlight the similarity between our masks (see Fig.1(i) and Fig. 3(d), for example) with the rotationally symmetric chiral patterns which are abundant in nature, from spiral galaxies (logarithmic spiral) in deep space to the seed arrangements in a sunflower head. Our analysis suggests such natural patterns may support vortex beams in nature and we are currently exploring bespoke vortex beams that can be experimentally produced from designs based on various naturally occurring patterns.

In conclusion, we have put forward and demonstrated the utility of a fundamental superposition and selection principle of OAM modes. This constitutes a new approach in the design of convertors for structured beam generation from plane waves, providing a versatile and robust alternative to the conventional CGH approach. A specific advantage of the complex vortex beams generated by the methods we have described is that it can produce complicated structured beams, such as concentric electron vortices and optical vortices, which carry different OAMs with different radii. These may be made use of in the manipulation of fluid-borne particles and also as a nano-optomechanical couette shear cell [36]. Our analysis provides a new understanding of the essential components of the OAM conversion process by rotationally symmetric wave-optical structures. It is applicable to the generation and analysis of OAM distributions of all scalar waves in axially symmetric systems.

This work is supported by EPSRC Grants No. EP/G070474/1 and No. EP/G070326/1 and COST Action MP0903. Y.J. Yang acknowledges the support by the National Natural Science Foundation of China under Grant Nos. 61205122 and 11474048.